\documentclass[runningheads]{llncs}
\usepackage[T1]{fontenc}
\usepackage{graphicx}
\usepackage{hyperref}
\usepackage{url}
\usepackage{xcolor}
\newcommand\man[1]{{#1}}

\newcommand\sysacronym{TROPIC}
\newcommand\sysname{Trustworthiness Rating of Online Publishers through online Interactions Calculation}

\begin{document}
\title{\sysacronym \ – \sysname}
\titlerunning{TROPIC – Trustworthiness Rating of Online
Publishers}
%
\author{Manuel Pratelli\inst{1, 2}\orcidID{0000-0002-9978-791X} \and
Fabio Saracco\inst{3,1,4}\orcidID{0000-0003-0812-5927} \and
Marinella Petrocchi\inst{2,1}\orcidID{0000-0003-0591-877X}}
\authorrunning{M. Pratelli, F. Saracco and M. Petrocchi}
%
\institute{IMT School for Advanced Studies Lucca, Lucca, Italy \and
CNR-IIT, Pisa, Italy
\and
`Enrico Fermi' Research Center (CREF), Rome, Italy\\
\and
CNR-IAC, Sesto Fiorentino, Italy
}
\maketitle              
\begin{abstract}
Existing methods for assessing the trustworthiness of news publishers face high costs and scalability issues. The tool presented in this paper supports the efforts of specialized organizations by providing a 
solution that, starting from an online discussion,  provides (i) trustworthiness ratings for previously unclassified news publishers and (ii) an interactive platform to guide annotation efforts and improve the robustness of the ratings. The system implements a novel framework for assessing the trustworthiness of online news publishers based on user interactions on social media platforms.

\end{abstract}
\section{Introduction}
The decline of traditional journalism, coupled with the increasing challenges of maintaining high standards of editorial quality, has raised concerns about the trustworthiness of online news  media. Specialized organizations~\cite{newsguard,mediabiasfactcheck,iffyindex,globaldisinformationindex,adfontesmedia}
provide valuable ratings that assess the credibility of digital news publishers. 
Although these ratings provide beneficial insights, they are based on the evaluation of resource-intensive and time-consuming criteria, such as the presence of biased and propagandist content in the news articles~\cite{bazmi2023multi,kim2019combating}. As a result, many online publishers remain unlabeled, creating a gap in coverage.
\\
\indent \textbf{Proposal:} 
Based on the work in \cite{pratelli24unveiling}, this demo paper presents \sysacronym -- \sysname \ (available at \url{https://tropic.iit.cnr.it}), an interactive tool to improve current methods for assessing the trustworthiness of an online publisher  and to avoid potential problems of limited coverage.
\sysacronym \ outputs include:

\begin{itemize}

    \item \textbf{The automatic classification of the level of trustworthiness of an online news publisher:} \sysacronym \ provides an assessment of the trustworthiness of unclassified online news publishers 
    by analyzing social media interactions between news producers and consumers.
The evaluation starts with the analysis of an online social discussion and leverages (i) a
\textit{base-knowledge} consisting of a subset of annotated news publishers, (ii) the concept of \textit{News Engagement Communities}, NECs for brevity, which are communities of news articles that received the most attention from discussion participants, 
 introduced in \cite{Pratelli2024,pratelli24unveiling} and 
(iii) the propensity of social users to share low-quality information. This approach provides (i) good accuracy-cost trade-off and (ii) good coverage of \textit{unclassified} and \textit{untrustworthy} news publishers \cite{pratelli24unveiling}.

    \item \textbf{An interactive guide for base-knowledge extension:} The robustness of the classification can be continuously improved by gradually extending the current \textit{base-knowledge}. The intuition is that an extended knowledge allows for more accurate social user profiling, and thus more accurate predictions about the trustworthiness of unannotated news publishers. To guide the expansion of the \textit{base-knowledge}, \sysacronym \ provides the end user with a set of functionalities that suggest the "best" next publisher to annotate. This guides where to focus the extension of the base knowledge, taking best advantage of the prediction system. 
    
\end{itemize}

\textbf{Context:} 
Manual methods of assessing the trustworthiness of online newspapers are costly, both in terms of the time required to perform the assessment and the cost of finding experienced reviewers. These barriers limit the reach of assessable newspapers. The low coverage also stems from the constant emergence of new, lesser-known online newspapers. \sysacronym\ aims to streamline the rating process and expand the reach of rated newspapers.
\\
\indent \textbf{Audience:} Designed for journalists, media professionals, and researchers evaluating the quality of  online information, \sysacronym \ provides estimates of the trustworthiness of online publishers.
The tool guides the user in the selection of online publishers to manually annotate, so that the subsequent automatic evaluation of news publishers not yet evaluated is effective in terms of prediction accuracy and coverage of the largest number of news publishers evaluated.
\section{System Overview}

\begin{figure}
\centering
\includegraphics[width=.8\textwidth]{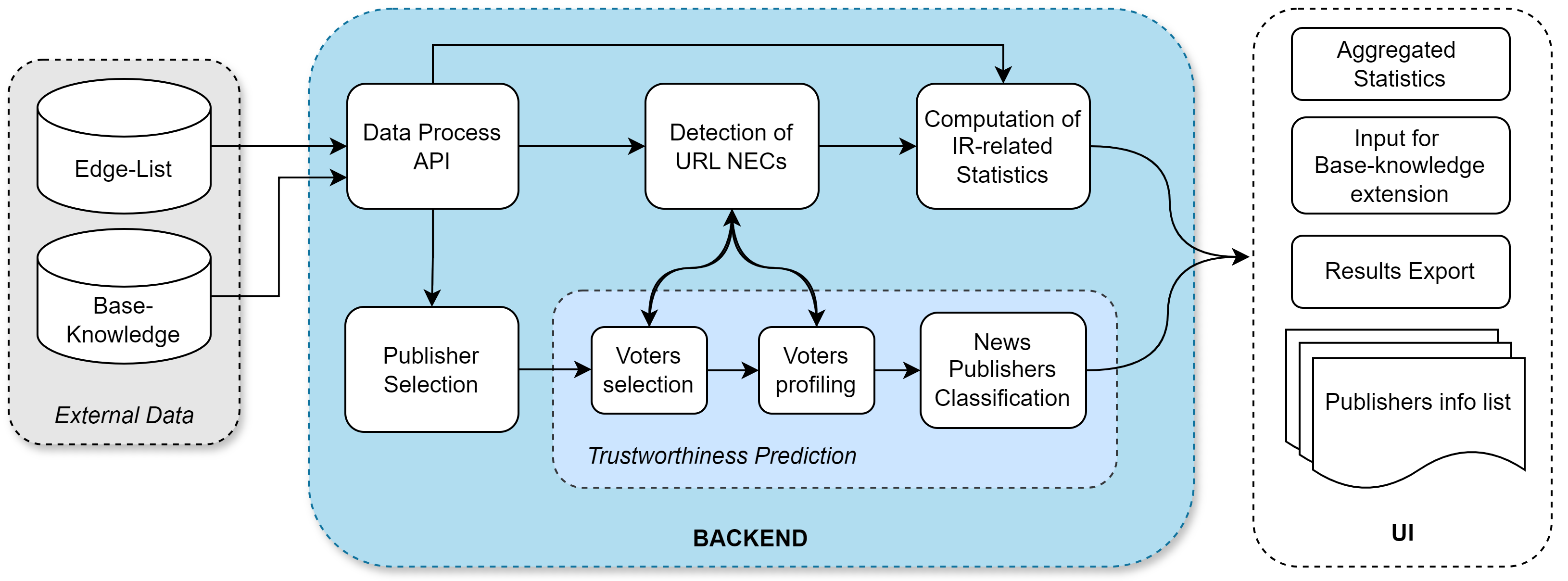}
\caption{System Overview} \label{fig:overview}
\end{figure}

\textbf{External Data:} 
Figure \ref{fig:overview} shows the tool's architecture. \sysacronym \ 
\man{takes as input an online discussion, in the format \textit{edge-list}, i.e., a list of URL - user ID pairs. The URL points to an online news story and is present in social posts, the user ID is the username of the social user who posted or shared that URL on the social platform.}
The tool is adaptable to data from any social media platform\footnote{URL preparation, such as resolving short URLs, is left to the \sysacronym \ user.
}. A second (optional) input for \sysacronym \ is the initial base-knowledge, which is a list of online publishers with associated trustworthiness scores (integers from 0 to 100). Since \sysacronym \ is designed to support organizations specializing in evaluating news publishers' trustworthiness, the initial knowledge base can consist of publishers already assessed by these organizations\footnote{\man{For example, NewsGuard \cite{newsguard} uses the 0-100 scale, the same as \sysacronym . In contrast, other organizations, such as Media Bias / Fact Check \cite{mediabiasfactcheck}, use descriptive labels and must be converted to numerical values to be used in \sysacronym .}
}.

\indent\textbf{Backend:} \man{As we can observe in Figure \ref{fig:overview}}, the \sysacronym \ backend consists of independent software components that work together to provide (i) a predicted trustworthiness score (with a corresponding confidence level) for each publisher not in the base-knowledge, and (ii) metrics to guide users on which publishers to annotate in case they want to manually expand the base-knowledge. The edge list is used both to build NECs communities (which we recall are communities of news URLs that have been very successful among online users in terms of engagement \cite{Pratelli2024,pratelli24unveiling}) and to calculate some quantity related to the online discussion we are studying (IR-related statistics). Once we have selected the publishers whose URLs have been shared in the online discussion, we move on to selecting a special group of users, called voters, and profiling them. After this profiling phase, these voters are asked to classify the publishers they interact with. 

\begin{figure}
\centering
\includegraphics[width=.45\textwidth]{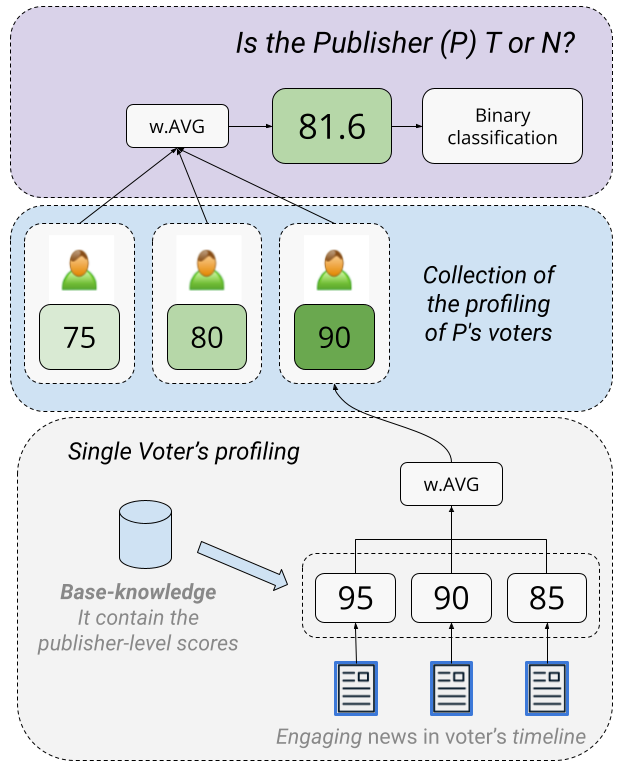}
\caption{Publisher Trustworthiness Classification} \label{fig:classification}
\end{figure}

In our current implementation:

\begin{enumerate} 
    \item We select as voters those users who have shared at least one engaging news (i.e., a news whose URL belongs to one of the NECs). 
    \item 
    For profiling voters - i.e., estimating each voter's propensity to share low-quality content - we follow the approach in \cite{pratelli24unveiling}, which consists of a two-step method to downscale the problem. First, instead of considering the complete set of URLs in a voter's timeline, we only select engaging URLs, specifically those belonging to NECs. \man{Second, instead of evaluating each individual online news article, we evaluate the trustworthiness of the URL by referencing the publisher-level score, if available, from the base knowledge (e.g., source-level scores of 95, 90, and 85, as shown in Figure \ref{fig:classification}, bottom). These publisher-level scores are aggregated using a weighted average, resulting in a single final score for each voter \man{(middle part in Figure \ref{fig:classification})}.
}
\end{enumerate}

Finally, to compute the trustworthiness score for an unclassified publisher, we select all voters who share content from that publisher, profile them, and aggregate their scores into a single (average) number that expresses the trustworthiness score of the considered publisher (Figure \ref{fig:classification}, \man{top}).
\begin{figure}
\centering
\includegraphics[width=\textwidth]{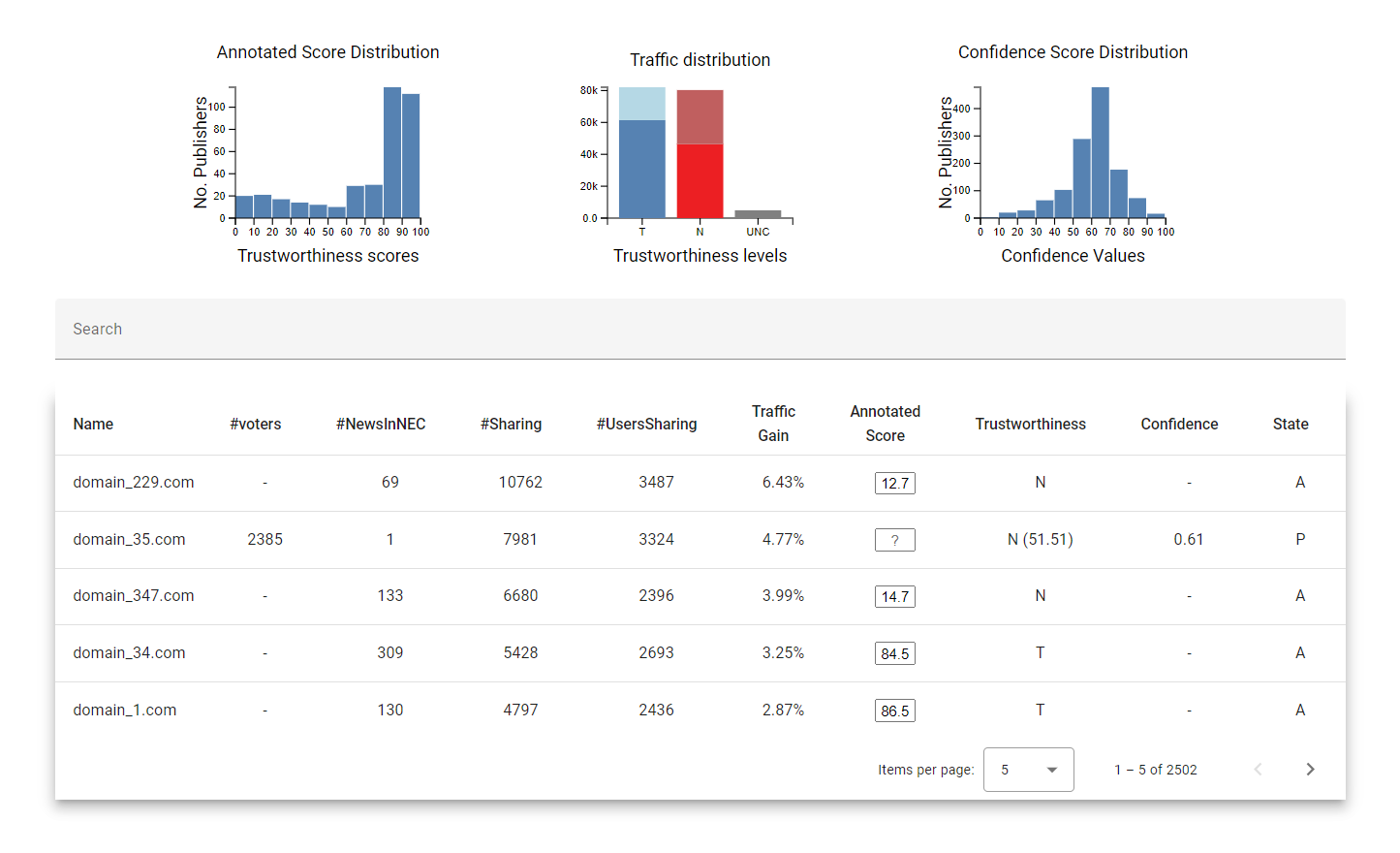}
\caption{The User Interface} \label{fig:ui}
\end{figure}
\\
\indent \textbf{User Interface:} Figure \ref{fig:ui} shows the User Interface (UI). The toolbar allows to upload the edge list (via the "File Upload" button) and, optionally, the \textit{base-knowledge}, and to initiate the computation via the "Process data" button. Calculation results are displayed in three plots and a table. The table lists each publisher's Name, IR-related statistics, an editable \man{Annotated} Score field (this allows the user to manually annotate other domains, if desired, to extend the base knowledge), a \man{Trustworthiness} column (showing both binary labels - trustworthy T/ untrustworthy N) and prediction scores, a Confidence column, and a State column (indicating if the rating comes from human annotation or from \sysacronym \ \man{- A stands for annotated, P stands for Predicted}). 
Users can sort the table by IR-related metrics, such as the number of voters or the number of news URLs shared within NECs. This way, manual annotations can be prioritized based on a metric of interest.
Above the table, plots provide information on the number of manually annotated publishers and the distribution of their scores (left plot); the number of publishers annotated by human experts (dark color at bottom) and those whose score are calculated by \sysacronym \ (light color at top) (middle plot); and the confidence level of the scores calculated by \sysacronym \ (right plot). This information guides the user to minimize the number of publishers that remain unclassified and to improve the confidence level of the predictions (by adding new manual annotations if necessary). The "Export Actual Knowledge" button allows to export the calculation results in CSV format.
\\
\indent \textbf{Implementation and Demonstration:} The user interface is built with Angular \cite{angular}, and the backend uses FastAPI \cite{fastapi}. NECs extraction employs the bicm Python library \cite{bicm}. Both UI and backend are containerized separately with Docker \cite{docker}.
For practical testing, we provide the tool in a DEMO version. In this configuration, the waiting times for calculating the NECs, which could be several minutes for a very large edge list, are pre-calculated. To utilize this setup, the end user is supplied with a DEMO edge list, which must be selected before pressing "Process Data". For completeness, a DEMO base-knowledge is also provided, featuring randomly generated trust scores. If end users want to upload their own edge list, the number of edges that can be uploaded in the demo version is limited to 50,000 entries.

\section{Conclusions}
In this demo paper, we presented an interactive web interface designed to streamline the annotation process for assessing the trustworthiness score of an online news publisher. \sysacronym \ predicts trustworthiness scores for publishers that have not yet been manually labeled, improving the coverage and reliability characterization of news domains traffic exchanged in an online discussion. 


\begin{credits}
\subsubsection{\ackname} This work is partially supported by SERICS (PE00000014) under the NRRP MUR program funded by the EU - \#NGEU. FS was partially supported by the project ``CODE – Coupling Opinion Dynamics with Epidemics'', funded under NRRP MUR Mission 4 ``Education and Research'' - Component C2 - Investment 1.1 - Next Generation EU ``Fund for National Research Program and Projects of Significant National Interest'' PRIN 2022 NRRP, grant code P2022AKRZ9.
\end{credits}

\newpage
%
%
%
\bibliographystyle{splncs04}
\bibliography{biblio}
\end{document}